\documentclass{elsart1p}
\usepackage{amssymb}

\newcommand{\be}{\begin{equation}}
\newcommand{\ee}{\end{equation}}
\newcommand{\bea}{\begin{eqnarray}}
\newcommand{\eea}{\end{eqnarray}}
\newcommand{\benn}{\begin{displaymath}}
\newcommand{\eenn}{\end{displaymath}}
\newcommand{\beann}{\begin{eqnarray*}}
\newcommand{\eeann}{\end{eqnarray*}}
\begin{document}

\begin{frontmatter}

\title{On pomeron loop effects in deep inelastic scattering}

\author{Wenchang Xiang}

\address{Fakult$\ddot{a}$t f$\ddot{u}$r Physik, Universit$\ddot{a}$t Bielefeld, D-33501 Bielefeld, Germany}

\begin{abstract}

Pomeron loop effects (gluon number fluctuations) on deep
inelastic scattering (DIS) in the fixed coupling case are studied.
We show that the description of the HERA data for the inclusive structure function $F_2(x,Q^2)$ for
$x\leq10^{-2}$ and $0.045\leq Q^2\leq 45GeV^2$ is improved once pomeron loop effects are included.
\end{abstract}

\end{frontmatter}

\section{Introduction}
\label{intro}
%

Recently, there has been a tremendous theoretical progress in understanding the high energy QCD evolution beyond the
mean field approximation, i.e beyond the BK-equation~\cite{Kovchegov:1999yj+X}. It has been understood how to include
the fluctuations of gluon numbers (Pomeron loops) in small-x evolution~\cite{Mueller:2004se, Mueller:2005ut, Iancu:2004es}.
A relation between high-energy QCD and statistical
physics was found which has clarified the physical picture and the way to deal with the dynamics beyond BK-equation.
After including fluctuations, the evolution becomes stochastic
and the physical amplitude, which one obtains by averaging over
all individual realizations.

\section{Mean field approximation}
\label{sec:mfa}
The Balitsky-Kovchegov (BK) equation~\cite{Kovchegov:1999yj+X}
gives the evolution with rapidity $Y=\ln(1/x)$ of the $T$-matrix
for a $q\bar{q}$ dipole scattering off a target which may be
another dipole, a hadron or a nucleus\cite{Xiang:2009prd},
\be
\frac{\partial T}{\partial Y} \sim \alpha_s[T - TT]
\label{Eq_Kovchegov}
\ee
%

One of the hallmarks of the BK-equation is the so called geometric scaling behavior of the scattering
amplitude,
\be
T(r, Y)=T(r^2Q_s^2(Y))
\label{geoscal}
\ee
where $r=|x_{\bot}-y_{\bot}|$ is the size of dipole and $Q_s(Y)$ is the saturation momentum defined such that
$T(r=1/Q_s, Y)$ be of order $1$. The $T$-matrix shows the geometric scaling behavior, i.e., it depends only on
a single quantity $r^2Q_s^2(Y)$ instead of depending on $r$ and $Y$ separately, which seems well supported by
HERA data\cite{Stasto:2000er,Iancu:2003ge}.

Another important result extracted from the BK-equation is the rapidity dependence of the saturation
scale,
\be
Q_s^2(Y)=Q_0^2\frac{\exp[\bar{\alpha}\chi'(\lambda_s)Y]}{[\bar{\alpha}Y]^{\frac{3}{2(1-\lambda_0)}}},
\label{sm}
\ee
where $\chi(\lambda)=2\psi(1)-\psi(\lambda)-\psi(1-\lambda)$ is the eigenvalue of the BFKL kernel.

Iancu, Itakura and Munier~\cite{Iancu:2003ge} have used the following ansatz for
the $T$-matrix,
\begin{equation}
T^{\mbox{\footnotesize IIM}}(r,Y) = \left\{ \begin{array}{r@{\quad,\quad}l}
1- \exp\left[ - a \ln^2(b\,r\,Q_s(x))\right] & r\,Q_s(x) > 2
\vspace*{0.5cm}\\
N_0 \left(\frac{r\,Q_s(x)}{2}\right)^{2\left(\lambda_s +
    \frac{\ln(2/r\,Q_s(x))}{\kappa\,\lambda\,Y}\right)} & r\,Q_s(x) <
  2  \ ,
\end{array} \right.
\label{eq:IIM}
\end{equation}
which obviously includes the features of the solution to the BK equation, to
compare the theory in the mean field approximation with the DIS data.  They
have used for the saturation momentum the leading $Y$-dependence of
Eq(\ref{sm}), $Q_s(x) = (x_0/x)^\lambda$, however, with $\lambda$ and
$x_0$ being fixed by fitting the DIS data. The constant $\kappa =
\chi^{\prime\prime}(\lambda_s)/\chi^{\prime}(\lambda_s) \approx 9.9$ is a LO
result coming from the BK-equation, $N_0$ is a constant around $0.5$ and $a$
and $b$ are determined by matching the two pieces in Eq.(\ref{eq:IIM}) at
$r\,Q_s =2$.

The ``BK-diffusion term'' in the IIM-ansatz~(\ref{eq:IIM}),
\be
\left(\frac{r\,Q_s(x)}{2}\right)^{2
  \frac{\ln(2/r\,Q_s(x))}{\kappa\,\lambda\,Y}} =
\exp\left[-\frac{\ln^2(4/r^2\,Q^2_s(x))}{2\,\kappa\,\lambda\,Y}\right] \ ,
\label{eq:dif:IIM}
\ee
does explicitly violate the geometric scaling behaviour. We
wish to emphasize here that, as also shown in~\cite{Iancu:2003ge},
this violation seems required in order to get an accurate
description of the DIS data. Without it, even allowing $\lambda_s$
to be an additional fitting parameter, one can not get a better
description of the DIS data. For further details on the importance
of the diffusion term see Ref.~\cite{Iancu:2003ge}.

\section{Beyond the mean field approximation}
\label{sec:bmfa}
%
%
\subsection{Event-by-event scattering amplitude}
\label{subsec:se}
In this paper we will use the Golec-Biernat, W$\ddot{u}$sthoff (GBW)~\cite{GolecBiernat:1998js} and Iancu,
Itakura, Munier (IIM)~\cite{Iancu:2003ge} models for the event-by-event amplitude.

The GBW model
\be T^{GBW}(r,x)=1-\exp\left[-\frac{r^2Q_s^2(x)}{4}\right] \ee is
one of the most simple models which shows geometric scaling,
$T(r,x)=T(r^2Q_s^2(x))$, and leads to a quite successful
description of the HERA data.

%
%
\begin{table}
\begin{center}
\begin{tabular}{r@{\quad}||c@{\quad}|c@{\quad}|c@{\quad}|c@{\quad}|c@{\quad}|c@{\quad}|}
model/parameters & \quad $\chi^2$ & \quad $\chi^2/\mbox{d.o.f}$ & \quad $x_0$
($\times 10^{-4}$) & \quad $\lambda$ &
\quad $R$(fm) & \quad $D$ \\ [0.5ex] \hline \hline
$T^{\mbox{\footnotesize GBW}}$ & \quad 266.22 & \quad 1.74 & \quad 4.11 & \quad 0.285 & \quad 0.594 &
\quad 0 \\ \hline
$\langle T^{\mbox{\footnotesize GBW}} \rangle$ & \quad 173.39 & \quad 1.14 & \quad 0.0546 & \quad
0.225 & \quad 0.712 & \quad 0.397 \\ \hline
\end{tabular}
\caption{GBW model: The parameters of the
  event-by-event and of the physical amplitude.}
\label{tab_GBW}
\end{center}
\end{table}
\begin{table}[htp]
\begin{center}
\begin{tabular}{r@{\quad}||c@{\quad}|c@{\quad}|c@{\quad}|c@{\quad}|c@{\quad}|c@{\quad}|}
model/parameters & \quad $\chi^2$ & \quad $\chi^2/\mbox{d.o.f}$ & \quad $x_0$
($\times 10^{-4}$) & \quad $\lambda$ &
\quad $R$(fm) & \quad $D$ \\ [0.5ex] \hline \hline
$T^{\mbox{\footnotesize IIM}}$ & \quad 150.45 & \quad 0.983 & \quad 0.5379 & \quad 0.252 & \quad 0.709 &
\quad 0 \\ \hline
$\langle T^{\mbox{\footnotesize IIM}} \rangle$ & \quad 122.62 & \quad 0.807 & \quad 0.0095 & \quad
0.198 & \quad 0.812 & \quad 0.325 \\ \hline
\end{tabular}
\caption{IIM model: The parameters of the
  event-by-event and of the physical amplitude.}
\label{tab_IIM}
\end{center}
\end{table}

\subsection{Physical scattering amplitude}
\label{ss:phys_amp}
Based on the relation between high-energy QCD evolution and reaction-diffusion
processes in statistical physics~\cite{Iancu:2004es, Brunet:2005bz}, the physical amplitude,
$\bar{T}(r,Y)$, is then given by averaging over all possible gluon number
realizations/events
\begin{equation}
\langle T((\rho-\rho_s(Y))\rangle = \int d\rho_s\ T(\rho-\rho_s(Y)) \
P(\rho_s(Y)-\langle\rho_s(Y)\rangle) \ \ ,
\label{av_gd}
\end{equation}
where have used $\rho=\ln(1/r^2Q_0^2)$ and $\rho_s(Y)=\ln(Q_s^2(Y)/Q_0^2)$, $T(\rho-\rho_s(Y))$
is the amplitude for the dipole scattering off a particular realization of
the evolved proton at $Y$ and the distribution of $\rho_s(Y)$ is a
Gaussian~\cite{Marquet:2006xm}:
\begin{equation}
P(\rho _{s})\simeq
\frac{1}{\sqrt{\pi \sigma ^{2}}}\exp \left[ -\frac{
\left( \rho _{s}-\langle \rho _{s}\rangle \right) ^{2}}{\sigma ^{2}}\right] \ .
\label{proba_gauss}
\end{equation}
The expectation value of the front position, $\langle \rho_s(Y)\rangle$,
increases with rapidity as $\langle \rho_s(Y)\rangle =\ln(Q^2_s(Y)/Q_0^2)$ at
high energy~\cite{Iancu:2004es}.
The dispersion of the front at high energy increases linearly with rapidity,
\begin{equation}
\sigma^2 = 2 \left[\langle \rho_s^2 \rangle- \langle \rho_s \rangle^2\right] = D \, Y
\end{equation}
\noindent where $D$ is the diffusion coefficient, whose value is known only for
$\alpha \to 0$ (asymptotic energy)~\cite{Brunet:2005bz,Mueller:2004se}.
Since the values of $D$ and the exponent $\lambda$ of the saturation scale,
$Q^2_s(x) =1\,\mbox{GeV}^2\, (x_0/x)^\lambda$, are
not known for finite energies, in what follows we will
treat them as free parameters.

\section{Numerical results}
\label{res}
We use the dipole picture~\cite{mueller90}, which is a factorization scheme for DIS, to study the proton structure function,
which leads to the following expressions for the $F_2$ structure function:
\bea
 F_2(x,Q^2) = \frac{Q^2}{4\pi^2\alpha_{em}}\ (\sigma_T(x,Q^2) +
 \sigma_L(x,Q^2)) \ \\ \quad \quad \sigma_{T,L}(x,Q^2) = \int \ dz\,d^2r \ |\psi_{T,L}(z,r,Q^2)|^2 \ \sigma_{dip}(x,r)
\label{eq:Fst}
\eea

We fit the ZEUS data~\cite{Breitweg:2000yn+X} for the $F_2$ in the kinematical range $x \leq 10^{-2}$ and $ 0.045 \,\mbox{GeV}^2 < Q^2 <
50\,\mbox{GeV}^2$.
The interesting ingredient for us in Eq.~(\ref{eq:Fst}) is the dipole-proton
cross section, $\sigma_{dip} = 2 \pi R^2 \ \langle T(r,x) \rangle$, with $2
\pi R^2$ being the outcome of the integration over the impact parameter.

Now, using the GBW model as an event-by-event amplitude, we
include the effect of gluon number fluctuations by averaging over
all events via Eq.~(\ref{av_gd}).  The resulting $\langle
T^{GBW}(r,x) \rangle$, which breaks the geometric scaling, leads
to a better description of the HERA data, as can
be seen from the comparison of the $\chi^2$ values in
Table~\ref{tab_GBW}. The large improvement after including
fluctuations seems to indicate that violations of geometric
scaling, and probably even gluon number fluctuations, are
implicated in the HERA data~\cite{Xiang:jhep}.

In the IIM case fluctuations do improve the description of the
HERA data, however not much, as can be seen from the comparable
$\chi^2/\mbox{d.o.f}$ values for $T^{\mbox{\footnotesize IIM}}$
and $\langle T^{\mbox{\footnotesize IIM}}\rangle$ in
Table~\ref{tab_IIM}. This is so because the IIM model does already
contain the geometric scaling violations via the BK-diffusion
term, $\ln(4/r^2Q_s^2)/\sqrt{2\kappa \lambda Y}$, and describes
accurately the HERA data, before including fluctuations. Since in
the case of the IIM model the fluctuations do not improve much the
description of the HERA data, one may conclude that the
BK-equation alone should describe the HERA data and that
fluctuations may be negligible in the energy range of the HERA data.
The intention of this work was to illustrate the possibility that
fluctuations may be present in the HERA
data~\cite{Xiang:jhep}.\\

\textbf{Acknowledgments}\\
The author gratefully acknowledges the collaboration with A.~I.~Shoshi
and M.~Kozlov and financial support by the Deutsche Forschungsgemeinschaft under contract Sh 92/2-1 and IRTG.

\end{document}